\documentclass[Afour,times, sagev]{sagej}

\usepackage{moreverb,url}

\usepackage[colorlinks,bookmarksopen,bookmarksnumbered,citecolor=red,urlcolor=red]{hyperref}

\def\volumeyear{2021}

\usepackage{cleveref}

\graphicspath{{figures/}{pictures/}{images/}{./}} 

\usepackage{microtype}                 
\PassOptionsToPackage{warn}{textcomp}  
\usepackage{textcomp}                  
\usepackage{mathptmx}                  
\usepackage{times}                     
\usepackage{tabu}                      
\usepackage{booktabs}                  
\usepackage[inline]{enumitem}                  
\usepackage{flushend}                  
\usepackage{tabularx}
\usepackage{makecell}

\usepackage{xcolor}
\usepackage{xspace}

\newcommand{\md}[1]{\textcolor{black}{#1}\xspace}

\newcommand{\TMP}{\emph{The Missing Path}\xspace}

\newcommand{\wsv}[1]{\raisebox{-.2\height}{\includegraphics[height=1em]{#1}}\,}

\begin{document}

\runninghead{Destandau and Fekete}

\title{The Missing Path: Analysing Incompleteness in Knowledge Graphs}

\author{Marie Destandau and Jean-Daniel Fekete}

\affiliation{Université Paris-Saclay, CNRS, Inria, LISN}

\corrauth{Marie Destandau, Université Paris-Saclay, CNRS, Inria, LISN}

\email{marie.destandau@gmail.com}

\begin{abstract}
\md{Knowledge Graphs (KG) allow to merge and connect heterogeneous data despite their differences; they are incomplete by design. Yet}, KG data producers need to ensure the best level of completeness, as far as possible. The difficulty is that they have no means to distinguish cases where incomplete entities could and should be fixed. We present a new visualisation tool: \TMP, to support them in identifying coherent subsets of entities that can be repaired. It relies on a map, grouping entities according to their \emph{incomplete} profile. The map is coordinated with histograms and stacked charts to support interactive exploration and analysis; the summary of a subset can be compared with the one of the full collection to reveal its distinctive features. We conduct an iterative design process and evaluation with 9 Wikidata contributors. Participants gain insights and find various strategies to identify coherent subsets to be fixed. 

\end{abstract}


\maketitle
\begin{figure*}
  	\frame{\includegraphics[width=\linewidth]{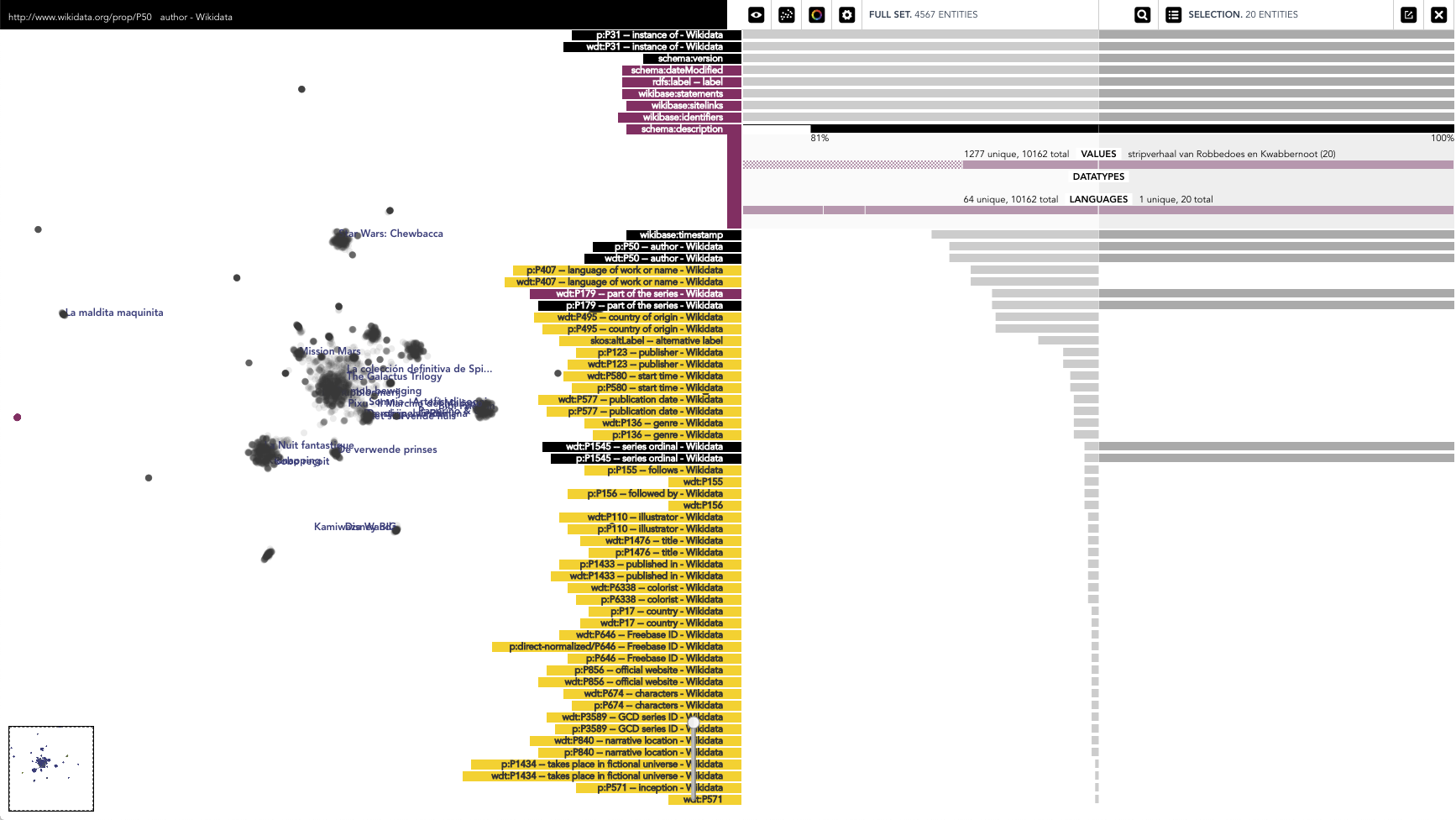}}
  	\caption{The map on the left shows the 4567 entities of type \texttt{wdt:Q1004 Comics} in Wikidata. The clusters appearing represent groups of entities that share the same missing paths.
  		If a collection were 100\% complete, there would be only one large cluster. 
  		The user has selected a small cluster of 20 entities on the left of the map; it is colored in dark pink.
  		On the left column, the histogram of paths completeness for the full collection can be compared with the histogram for the selected subset on the right.
  		Each row represents a path as a grey bar; its length is mapped to its percentage of completeness.
  		The left part of a row is colored in yellow if the path is missing in the selected subset and in dark pink if there is a significant difference between the full collection and the subset summaries.
	}
	\label{fig:teaser}
\end{figure*}
\section{Introduction}

\md{Knowledge Graphs (KG) allow to merge and connect heterogeneous data despite their differences, and this flexibility is key to their success. As they are incomplete by design, the drawback is that entities in a collection can have heterogeneous descriptions, potentially producing unreliable query results. Data producers, people, and organization producing KG data, still need to ensure, as far as possible, the best level of completeness. Completeness is regarded as an essential criterion in most quality methodologies for RDF data, the most used framework to describe KGs}~\cite{mendes2012sieve, harth2012completeness}. 

\md {The difficulty is that they have no means to distinguish cases where incomplete entities can and should be fixed. Let us consider the example of a publishing company building a KG with the books they publish and related books, such as those which inspired their books or are quoted in them. The graph merges data coming from several databases in the company, regularly enriched with information gathered from external sources such as Wikidata~\cite{MullerBirn2015} or Geonames~\cite{geonames}. Connecting their data in a KG enables them to power recommendation algorithms, to running analysis of the sales, and so on. Sharing this KG also allows researchers in humanities to analyse the books, and libraries and resellers to reuse the metadata. However, if the data are incomplete, such applications may lead to erroneous results, such as wrong decisions based on the incomplete analysis. While allowing incomplete information makes it possible to merge the various sources, some of the missing data might be fixed, but the various strata of data that were added and modified at different points in time make it difficult to identify them.}

Available tools and methods assess a completeness rate to each property in a collection and produce flat lists of all entities missing each property. Once assessed that, for instance, the publication date is missing for 11\% of the books (1346 books), the data manager has to inspect one by one a list of 1346 entities. After uncountable hours and thousands of clicks, she will maybe find out that certain issues were recurrent, and that she could have fixed them \md{in bulk}. She might realise that more than a hundred books came from the same original database describing the related books and that the date was actually not missing in the original data, but happened to be an uncertain date, expressed by a year, followed by a question mark (e.g.\ `1943?'). She might also notice that several of them had been published during war periods, while another significant part of them had been published clandestinely. Finding what those subsets had in common at the start would have given her a very useful hint: it was very unlikely that she would find more precise information by looking for the date in external data sources; she could have spared hours of unsuccessful research. Other subsets of interest could include books planned for publication, which can only be fixed later, when the date is known; or all books from a specific source, pointing to a bug in the transformation process, in which case she would rather fix the bug and run the transform script again rather than fixing entities one after another; and so on.
 She might also never notice those facts, as it is very difficult to find the coherence of scattered items when inspecting them in random order, especially if there are many meaningful subsets. 

Our tool, \emph{The Missing Path}, aims at addressing this issue. The map, grouping entities according to their incomplete profile, \md{lets users identify consistent subsets}. Comparing a specific subset with the full collection reveals its distinctive features, giving useful hints to understand the cause of incompleteness, and fix entities \md{in bulk}, saving significant time. \emph{The Missing Path} considers the completeness not only of direct properties (e.g.\ the \md{publisher}  of a book) but also of indirect properties  (e.g.\ the location of the \md{publisher} of a book), also called \emph{paths} of properties.
The novelty of our approach is 1) to use a map to identify structural similarity of entities in a KG, and 2) to support comparative analysis of the distributions of values at the end of paths of properties in a KG. 

Our contributions include:
\begin{itemize}[nosep]
    \item A method to transform a collection of entities into a map based on their incompleteness;
    \item A visualization tool called \emph{The Missing Path}, based on 3 coordinated views focused on the completeness, to support iterative exploratory analysis;
    \item A description of the iterative design process we used to improve and validate our approach while working with 9 Wikidata contributors.
\end{itemize}

We first introduce the basics of RDF and discuss related work regarding the evaluation and the visualisation of completeness. Then, we present the tool; we describe how path-based summaries are extracted and computed, we explain the design rationale and the main parts of the interface, and we illustrate it with a use case featuring a fictional Wikidata contributor. Eventually, we relate the iterative design process we used to improve and validate our approach while working with nine Wikidata contributors, following a methodology inspired by the ``Multi-dimensional In-Depth Long-term Case Studies'' (MILCS) of Shneiderman \& Plaisant~\cite{Shneiderman2006}.
The tool is available as open-source at: 
\href{https://gitlab.inria.fr/mdestand/the-missing-path/}{gitlab.inria.fr/mdestand/the-missing-path} and can be run online at: \href{https://missingpath.lri.fr/}{missingpath.lri.fr}. 

\section{Background and Related Work}
We introduce RDF and we discuss related work regarding the assessment and visualisation of their completeness.

\subsection{Introduction to RDF data}
RDF data are graph data; \md{their power relies on their structure: they} are made of low-level statements, named triples, that can be chained to answer complex queries, possibly over several data sources. \texttt{example:AuthorA schema:author example:BookB} is a triple, stating that Author A is the author of Book B. Triples are composed of a \emph{subject}, a \emph{predicate} and an \emph{object}. \emph{Subjects} are always \emph{entities}, represented by URIs. For readability, URIs can be prefixed: \texttt{example:AuthorA} stands for \texttt{<http://www.example/AuthorA>}. \emph{Predicates} ---also named \emph{properties}---also URIs; they follow rules defined in domain-specific models named \emph{ontologies}. \texttt{Schema.org} is an ontology specialised in the description of web pages, and \texttt{Schema:author} is one of the properties defined in it. \emph{Objects} can be \emph{entities} or \emph{literals}. When an object is an entity, it is possible to chain statements, for instance: Author A is the author of Book B, Book B's publisher is Editor C, Editor C's location is City D, City D's name is 'Paris'. The chaining stops when the object is a literal, like \texttt{'Paris'}, since a literal cannot be the subject of another triple. A chain of predicates is named a \emph{path} in the graph. 

The RDF framework is very flexible and allows each entity to be described with different properties. However, to make their data meaningful and usable, data producers need to ensure a minimum of homogeneity.

\begin{figure}[h]
  \frame{\includegraphics[width=\columnwidth]{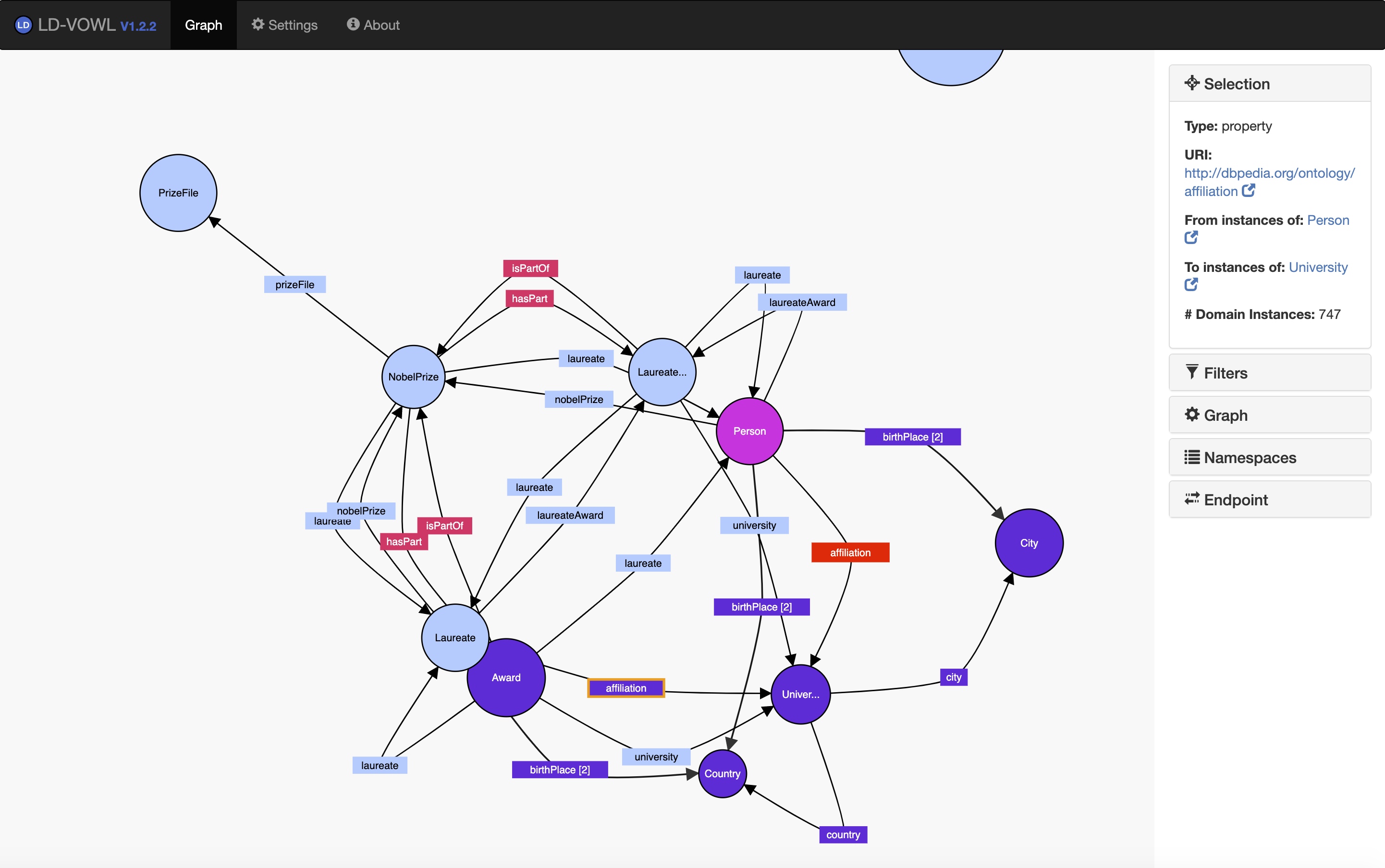}}
  	\caption{\md{Screenshot of LD-VOWL, taken on 2020-12-12 at \href{http://vowl.visualdataweb.org/ldvowl/}{vowl.visualdataweb.org/ldvowl}. The user has selected the property `affiliation' (in red) and can see in the top right panel that it is used 747 times. To know the rate of completeness of this property relative to the class Person, she needs to select the node Person, read in the panel that there are 910 instances, and compute that 747/910*100 = 82\% of the persons have an affiliation.
  	}}
	\label{fig:rw1}
\end{figure}
\begin{figure}[h]
  \frame{\includegraphics[width=\columnwidth]{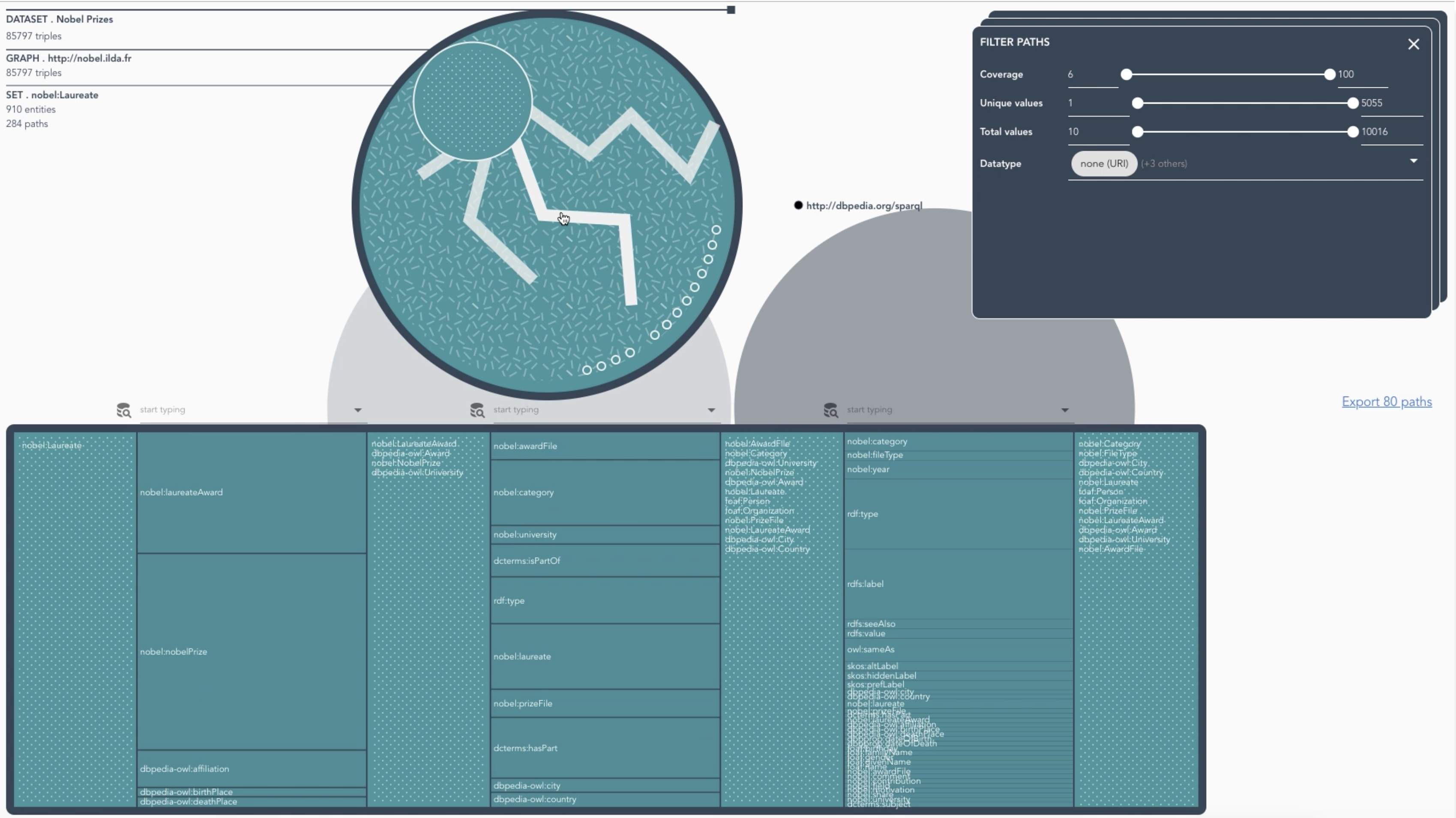}}
  	\caption{\md{Screenshot of Path Outlines, taken on 2020-12-05 at \href{http://spf.lri.fr/}{spf.lri.fr}. The user can browse the paths for a collection, filtering them on their completeness rate (among other metrics), and inspect the completeness rate of each path.}}
	\label{fig:rw3}
\end{figure}

\subsection{Completeness in RDF}
Though the definition of quality in RDF can have many acceptations, most work on the topic mention completeness as important criteria~\cite{mendes2012sieve, BIZER20091,radulovic2018comprehensive, zaveri2016quality, BenEllefi_2018}. The rate of completeness of a property is the percentage of entities in a given set described by this property. 
The set of entities can be the dataset or a subset. Technically speaking, approaches considering the dataset~\cite{auer2012lodstats} give the most accurate overview. However, from an editorial point of view, and except for some very generic properties, like \texttt{rdfs:label}, that might apply to any entity in a dataset; it is more reasonable to expect homogeneous descriptions for groups of entities that are similar, also named collections of entities. Issa et al.~\cite{issa2019revealing} use the class of resources as similarity criteria, for instance \texttt{schema:Person}, \texttt{schema:Organization}, or \texttt{schema:Place}, \md{and display the result as a UML class diagram. They do not support the evaluation of the completeness of paths of properties. A typical use case would be to evaluate the percentage of authors whose place of birth has geocoordinates, to know if plotting a map would give a representative overview of authors.}
\md{Using a node-link diagram to lay out a summary graph of the dataset allows to read paths of properties~\cite{Troullinou2018, weise2016ld}, as displayed in \autoref{fig:rw1}, with the limitation that the counts are given as absolute counts for each selected element. The user has to compute the rate himself for single properties, and cannot access it for paths of properties. To address this limitation, Path Outlines lets users browse paths following their completeness rate and other metrics~\cite{destandau2020}, as displayed in \autoref{fig:rw3}.}

\begin{figure}[h]
  \frame{\includegraphics[width=\columnwidth]{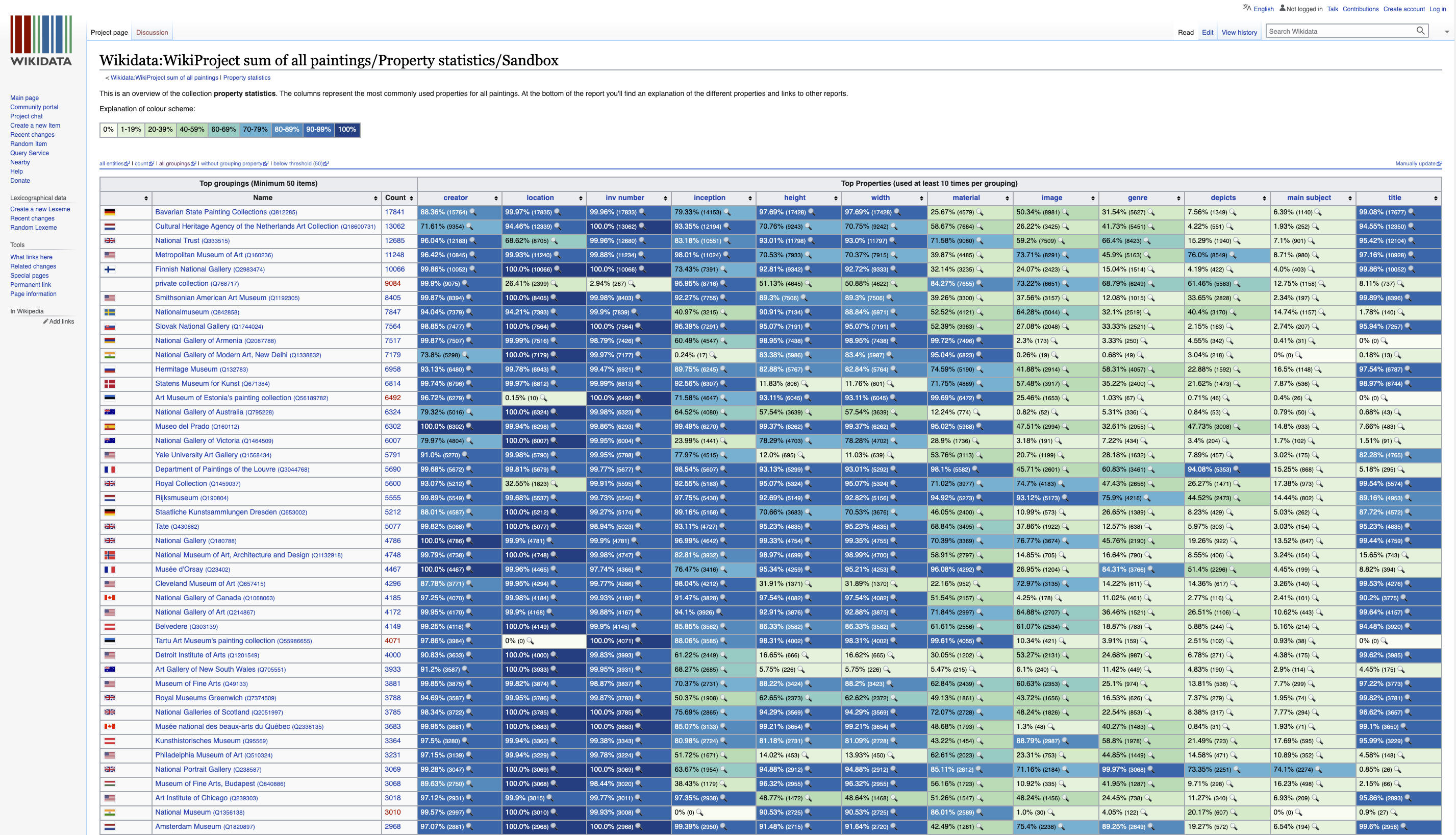}}
  	\caption{\md{Screenshot of Integraality for Wikidata,  taken on 2020-12-12 at \href{https://www.wikidata.org/wiki/Wikidata:WikiProject_sum_of_all_paintings/Property_statistics/Sandbox}{wikidata.org/wiki/Wikidata:WikiProject\_sum\_of \_all\_paintings/Property\_statistics/Sandbox}. The color scale helps users compare the completeness rate in the different groups. However, as the table scrolls over more than 5 screen heights, it is actually difficult to read and use. }}
	\label{fig:rw2}
\end{figure}

\md{However, considering the rate of completeness of a property or a path of properties relative to the full collection might not always be enough to help data producers fix their datasets.
In RDF, meaningful aggregation can also be achieved through the values of a property. For instance, entities in the collection \texttt{schema:Person} could be considered regarding their profession, encoded in the value of \texttt{schema:hasOccupation}. This allows identifying smaller subsets with similar profiles and needs. Integraality~\cite{InteGraality} lets users select a property to define subsets in the collection, and then evaluate the completeness of other properties relative to those subsets, as displayed in \autoref{fig:rw2}. The limit is that the table can be huge and thus difficult to read and use and that one single property might not produce useful groups to analyse the completeness of all properties. PRO-WD~\cite{wisesa2019wikidata} supports crossing several properties, but produces a grid of charts that is very difficult to interpret.}

\md{Our approach, instead of starting from the values at the end of properties to define consistent groups, identifies clusters of entities with a similar structure, in order to automatically reveal meaningful contexts concerning the properties over which completeness is evaluated.}

\begin{figure*}
  \frame{\includegraphics[width=9.7em]{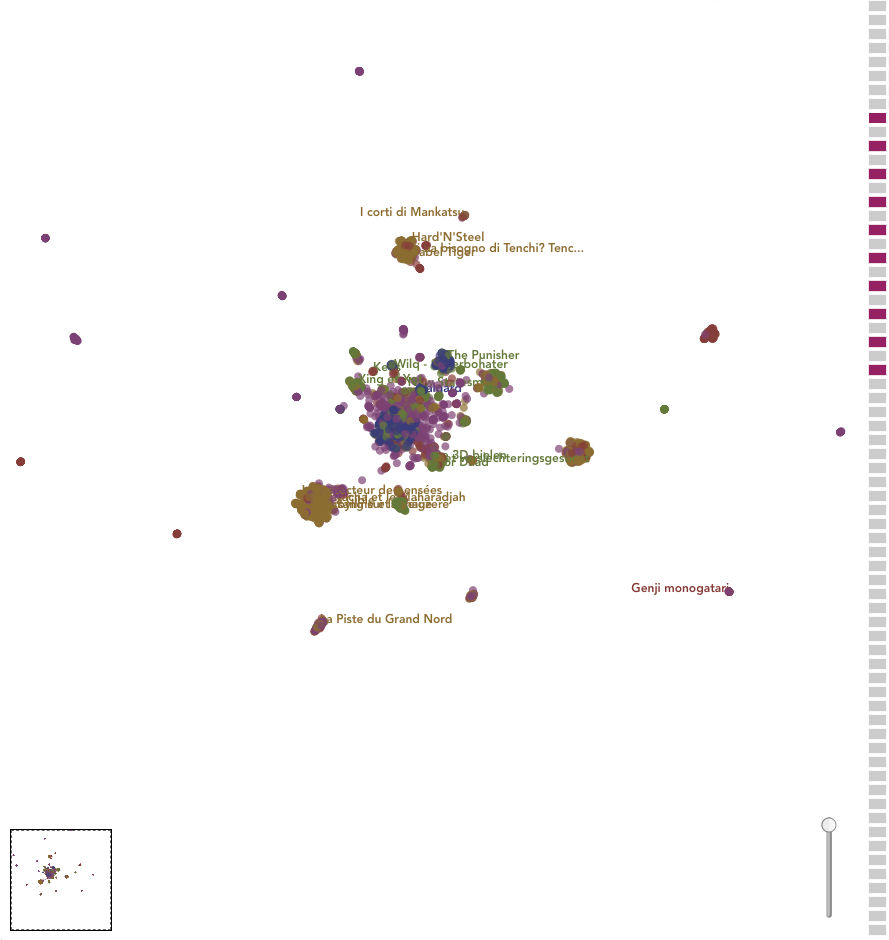}}
  \frame{\includegraphics[width=9.7em]{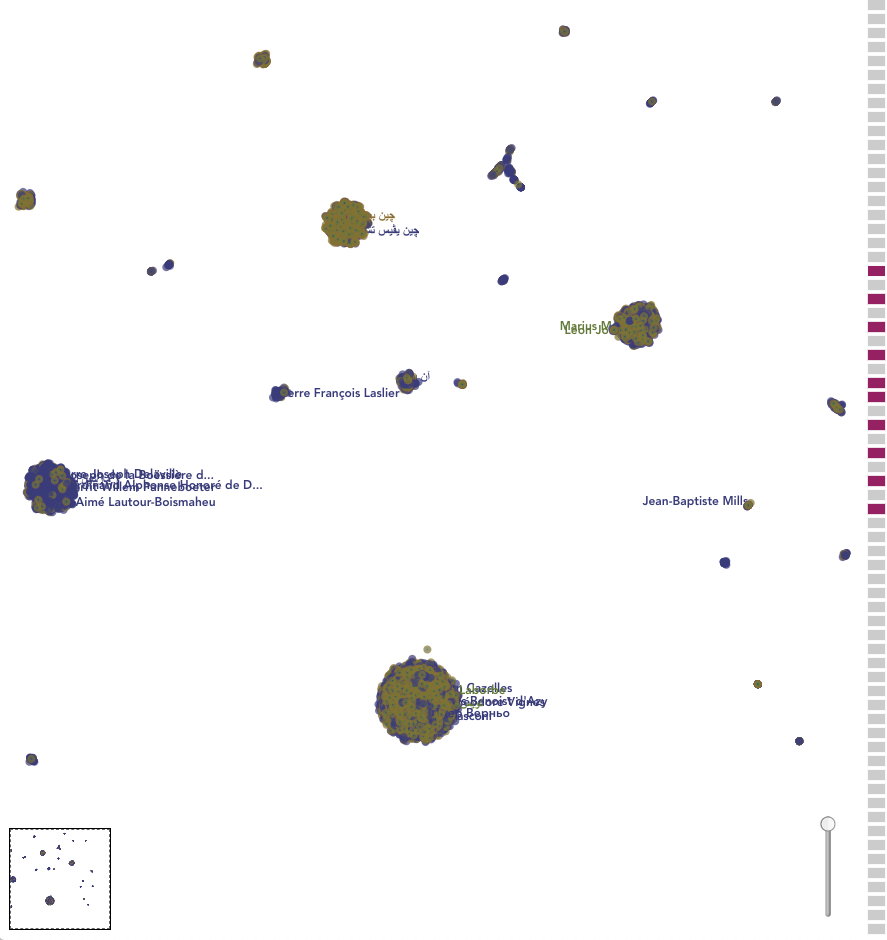}}
  \frame{\includegraphics[width=9.7em]{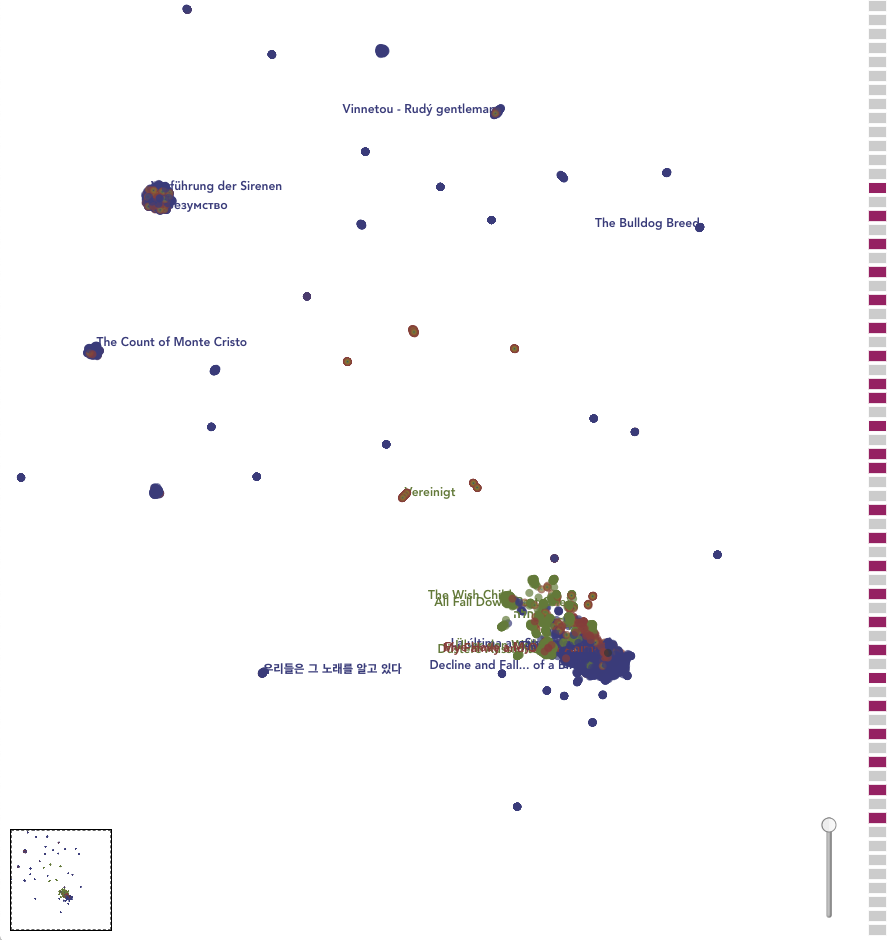}}
  \frame{\includegraphics[width=9.7em]{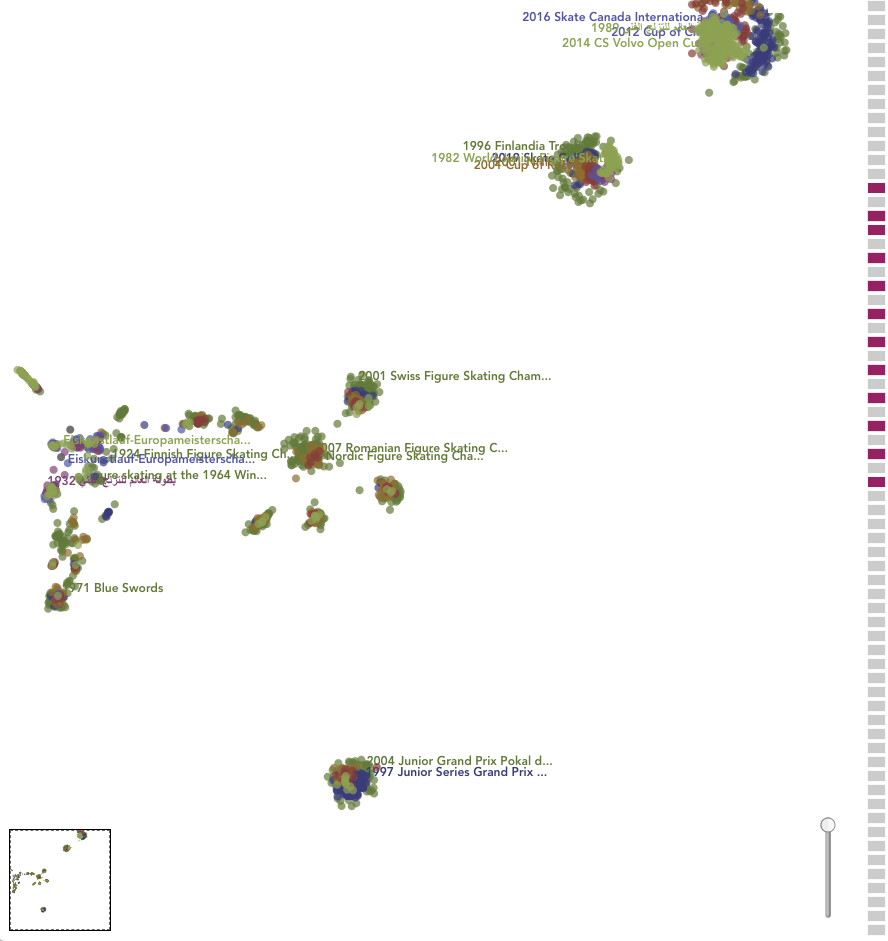}}
  \frame{\includegraphics[width=9.7em]{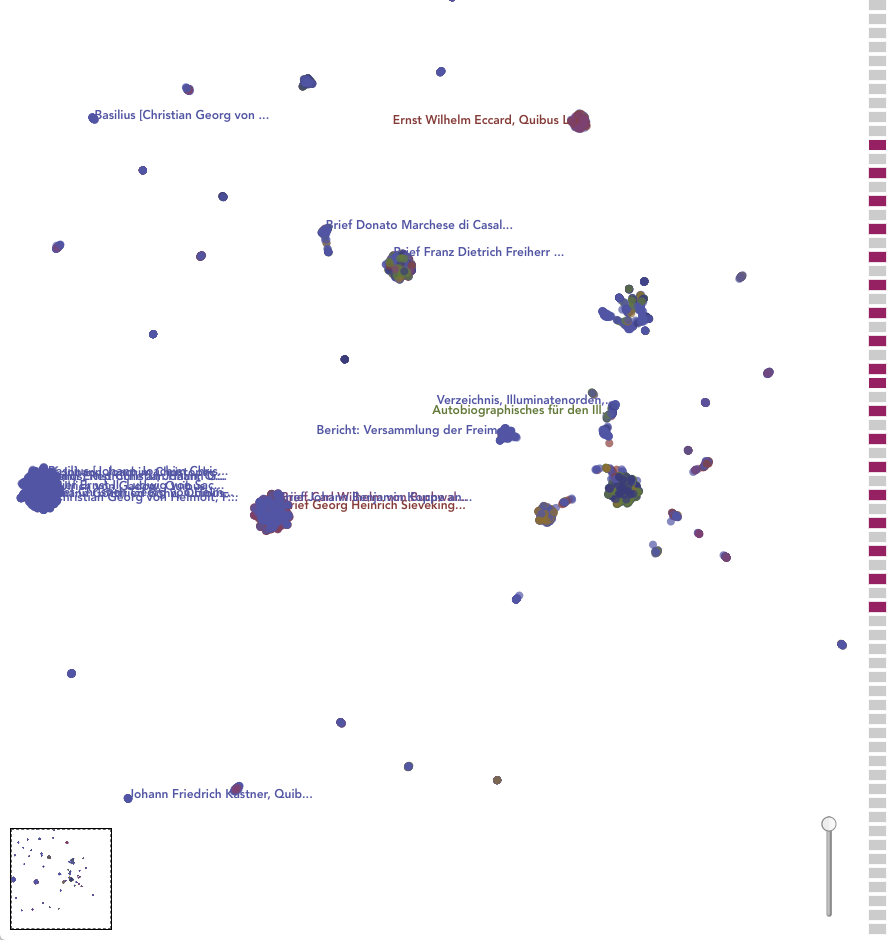}}
  	\caption{
  Collections C1, C2, C3, C4 and C6 (see~\autoref{tab:collection}). The number of clusters, their size and distribution provide a visual footprint of the shape of a collection, relative to the set of paths selected to produce the map (highlighted in pink on the right side of each thumbnail).}
	\label{fig:maps}
\end{figure*}

\begin{figure}
  \frame{\includegraphics[width=4.6em]{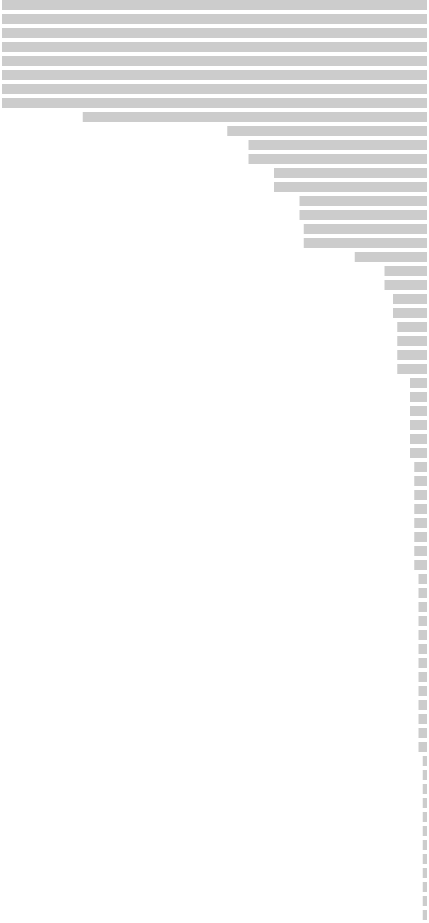}}
  \frame{\includegraphics[width=4.6em]{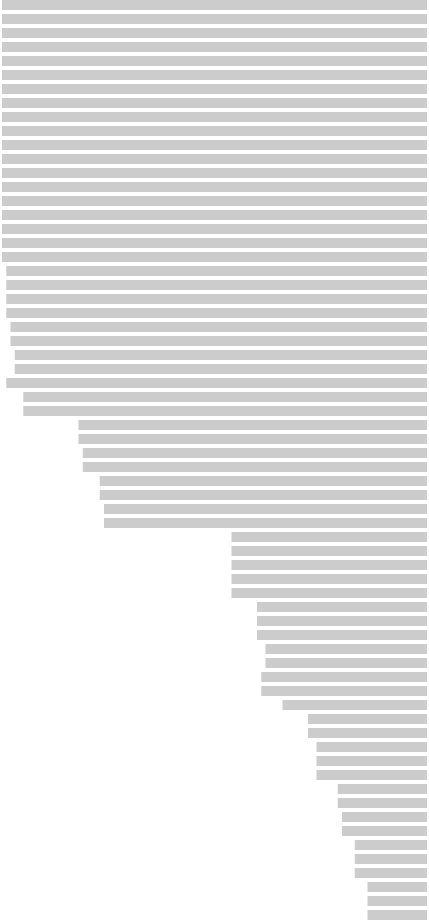}}
  \frame{\includegraphics[width=4.6em]{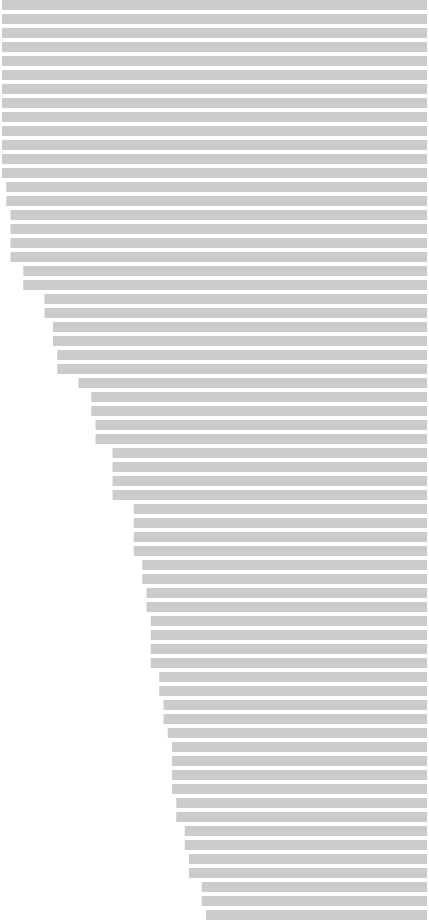}}
  \frame{\includegraphics[width=4.6em]{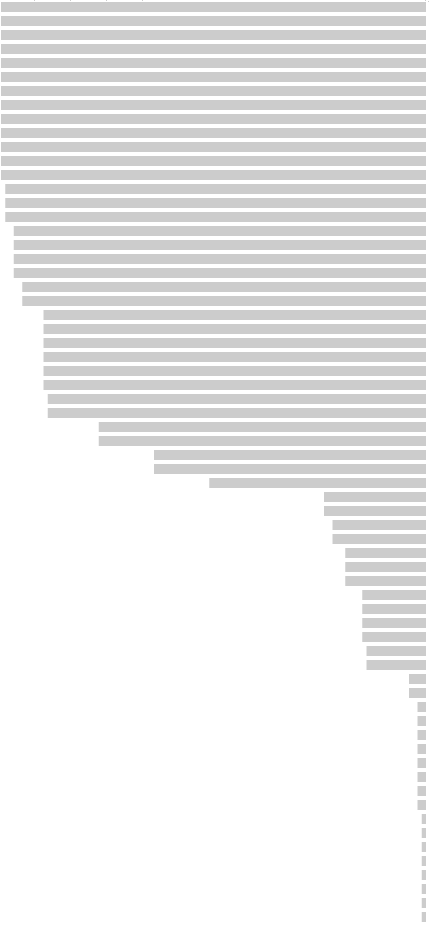}}
  \frame{\includegraphics[width=4.6em]{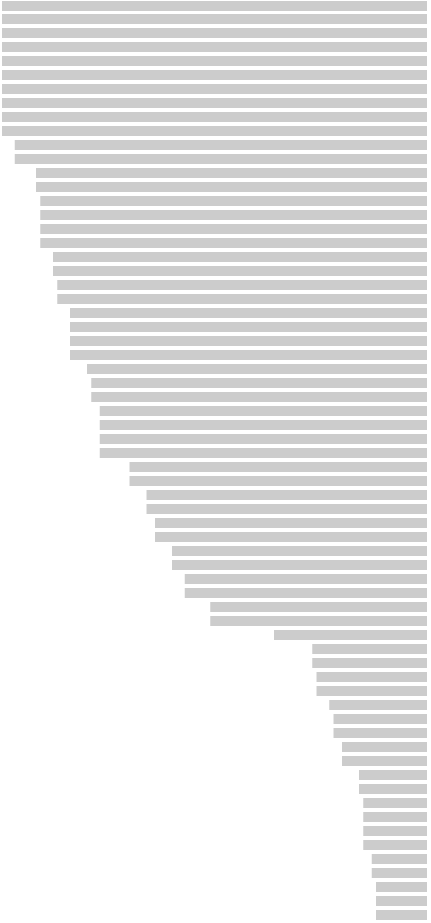}}
  	\caption{
  Collections C1, C2, C3, C4 and C6 (see~\autoref{tab:collection}). Histogram on the frontpage: the steepness of the curve gives a visual footprint of the completeness of the most complete paths in the collection. Scrolling down allows seeing all paths. C1 is our demo collection: it was not curated as a wikiproject, so very few paths are fully complete, and there is a sharp decrease with a long tail of paths with a low rate of completeness. C2 is maintained by an active team of 10 contributors: a large number of paths is complete. C3 is a catalog of films curated before it was imported: is more balanced. C4 has been created and curated over a short time mostly by one contributor. C6 is a starting project mixing sets of data that were curated separately.}
	\label{fig:charts}
\end{figure}

\section{\md{Data Representation and Processing}}
\md{The originality of our approach relies on the data representation. We build on the concept of \emph{semantic paths} to summarise the description of a collection, and we use the semantic paths as indexes for vector embeddings to compute a map of completeness as well as detailed summaries.}

\subsection{Paths summaries}
We build on the concept of \emph{semantic paths} to describe a given collection; they encode aggregate information relative to chains of triples. In the article introducing them~\cite{destandauSPaths2020}, their description is limited to the counts of unique and total values at the end of the chain. \md{We use vector embeddings to extend them and offer a} detailed summary of their distribution.
Our API takes as parameters \emph{the URI of a SPARQL endpoint}, \emph{a similarity criterion for the collection}, \emph{a maximum depth for the chains of properties} to analyse. We extract RDF data to process them into a matrix.
We first retrieve all the path patterns---the combinations of chains of properties that will be analysed---up to the max depth, and their completeness rate, as described in \cite{destandau2020}. The list is ordered by completeness, starting with the most complete path, and stored in a file. We assign an auto-incremented index as an identifier to each path.

\subsection{Retrieval of the entities}
Then we fetch the URIs of all entities in the collection. \md{A specific issue with SPARQL endpoints is that an endpoint cannot return more than a given number of lines as a result. This \texttt{quota} is usually set to 10,000 by default. Unlike SQL databases, there is no guarantee to retrieve all the results repeating the same query using the LIMIT, START, and ORDER BY commands. We use the \emph{semantic paths} to find a path to formulate several queries so that each query will retrieve less than \texttt{quota} entities. We initially set a \texttt{maxUniqueValues} variable to 30 in order to keep the number of queries reasonable. We start by checking the best-represented path with less than \texttt{maxUniqueValues} values at the end, and we retrieve the unique values at the end of this path, and the count of entities associated with each. If the highest count is lower than \texttt{quota}, and the number of entities not represented by this path is also lower than \texttt{quota}, we use this path to retrieve the entities: for each value, a query fetches all the entities having this value at the end of the path; then the last query fetches all entities not described with this path. We merge and deduplicate all the entities retrieved. We assign an auto-incremented index as an identifier to each entity. The list is stored in a file. Otherwise, if the path does not match the requirement, we consider the next path. If none of the paths meets the requirements, we increase \texttt{maxUniqueValues} and check the list of paths again.}

\subsection{\md{Values as vector embeddings}}
Each entity is described as a vector, where each column is a path. The value is either 'null' or a list of descriptors, structured as follows: \texttt{[values, datatypes, languages]}. Each element is itself a list, to account for multiple values, since cardinality is not constrained in RDF. For instance, a cell describing the label of an entity with 2 labels could contain: \texttt{[['À la recherche du temps perdu', 'In Search of Lost Time'], null, ['fr', 'en']]}. The datatype descriptors are filled only if they are expressed in the data. A cell describing the publication date of an entity could contain: \texttt{[[1998], ['xsd:dateTime'], null]}. \md{The vector is stored in a dictionary, associated with the URI of the entity.}

\subsection{\md{The completeness matrix}}
\md{The matrix of completeness is created from those vectors, each row is an entity.
The values are transformed as follows: 'null' becomes 1, meaning that a path is missing, and a list becomes 0.
Then, we project the vectors in 2 dimensions, to be later used as coordinates on a map.}

Dimensional reduction techniques~\cite{Nonato2018} allow computing clusters and lay them out on a map. They usually group entities according to the values of their core attributes (for instance, the topics of a set of books and their publication date), to have items with similar descriptions grouped together~\cite{Zaveri2013, hogan2010some}. We fill the vector with the structure of the description; we consider entities as similar if they are described by the same paths of properties, even if the values at their ends are not the same, to identify groups of entities missing such information.
Among the large number of dimensionality reduction techniques available~\cite{Nonato2018}, we opted for UMAP~\cite{umap}; this flexible method accepts both simple or sparse vectors---as we knew that the number of paths to consider, that is, the number of dimensions in a vector, could vary significantly across datasets---, and is fast and efficient for clustering.
We use UMAP with the dissimilarity function \emph{Russel-Rao} from the Scipy library~\cite{scipy}. This function computes a dissimilarity that takes into account the indices of the Boolean values in the vector---as opposed to a Jaccard function, for instance. As a result, items that form clusters on the map are those missing the exact same set of paths---while Jaccard would have grouped entities missing the same number of paths. To our knowledge, using maps to identify similarities in KGs is a novel approach.

\subsection{\md{Advanced summaries}}
\md{To produce the summaries, we construct a matrix with all the vectors, and we transform it into a table (a Pandas DataFrame) to compute the summaries with Python.}

\begin{figure}
  	\frame{\includegraphics[width=\columnwidth]{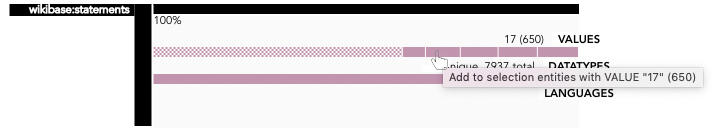}}    
  	\frame{\includegraphics[width=\columnwidth]{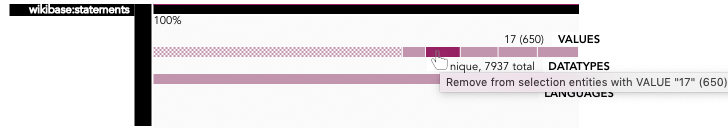}}
  	\caption{The user can click on an element of the summary to add it to the selection (top). Once added, it becomes dark pink, and clicking again will remove it (bottom).}
  	\label{fig:selectionsummary}
\end{figure}

The summaries are based on unique values. All values with a number of occurrences lower than 5\% of the total number of values are merged in an `other' bucket to keep the overview readable. The graphical elements can be used to select entities by clicking on them, as displayed in \autoref{fig:selectionsummary}.  The `other' bucket can also be used as a selector, and its values will be detailed as the selection narrows down.

To detect statistically significant differences, the system uses the distributions of the values at the end of a path for the subset, and for the full set, as displayed in the summaries, including the `other' aggregate. It normalises them and compares them against each other, performing a Kolmogorov-Smirnov test, using the \verb|scipy.stats.ks\_2samp| Scipy function. It then repeats this operation with the summaries of the datatypes and languages. If there appears to be a significant difference (p-value $< 0.1$) in either values, datatypes, or languages, the path is colored in pink.

\section{\md{User Interface}}
\md{This new data representation allows us to design an interface to analyse the incompleteness of subsets in RDF data~(\autoref{fig:teaser}). We will present the design rationale and detail the main parts of the interface: the map, the histograms with embedded stacked charts, and the selection bar.}

\subsection{Design rationale}
\md{To support the identification and analysis of subsets of entities relative to the completeness of their paths, The Missing Path coordinates an entity-centric visualisation, the map, with a path-centric visualisation, the histogram.
The map represents all the entities in the collection and allows to situate a selected subset through explicit color encoding (in pink) of selected items. The path summaries describe the full collection and the selection and are laid out in mirror.} The combination of \textit{superposition} (on the map) and \textit{juxtaposition} (in mirror) allows the effective support of comparison~\cite{Gleicher2011}. The tight integration of statistics and visualisation is known to support explorative data analysis~\cite{perer2008integrating}, helping users to make sense of the data.

There are two ways to select a subset of items sharing a similar structure: selecting a cluster on the map or using a combination of graphical elements in the summarised distributions to express logical constraints, e.g.~\emph{all items missing pathA and pathB, but not missing pathC}. The map is intended to guide users in their discovery, while the summarised distributions help them to refine a selection, or to fully express their own constraints to pursue their ideas when new ideas come to them~\cite{Perer2008SYF}.

\subsection{2D map of entities}
On the left part of the screen, the map (\autoref{fig:teaser}) displays clusters of items with similar incomplete profiles, offering an overview of the entities in the collection and allowing to select the clusters. It supports the following tasks:
\begin{itemize}[nosep]
\item see the homogeneity of the collection, regarding the completeness of the paths selected to compute the projection;
\item select subsets, through \md{precomputed} groups or using the interactive lasso; and
\item identify entities that are selected.
\end{itemize}

\subsubsection{Overview of the completeness}
\autoref{fig:maps} shows that different collections have different footprints. If a collection were 100\% complete, there would be only one large cluster. The number of clusters, their size, and distribution, form a visual footprint giving the shape of a collection relative to the set of properties selected to produce the map. Users can modify the list of paths taken into account to build the vector with the projection button \wsv{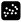} and recompute the map. Our Python API, based on UMAP-learn \cite{umap-learn}, takes a few to 30 seconds to recompute the map for the collections in \autoref{tab:collection}.

For instance, selecting only 2 properties, P1 and P2, to compute a map, could result in 4 clusters: entities missing both P1 and P2 properties, entities missing none of them, entities missing only P1 property, and entities missing only P2 property. The interest does not lie in the systematic enumeration of all combinations (in which case a table would be as efficient as a map). In reality, when more properties are taken into account, not all combinations happen, some are very frequent, and other concern only a few entities, and the map reveals unexpected clusters serving as entry points to explore a collection. Inspecting the profile of a cluster often reveals other similarities, that may relate to the provenance, the history, or the contributor.

\subsubsection{Colors}
\label{sec:map:color}
While the position of the entities is based on missing information, their color is linked to the content of present information. Paths for which the summary of values has more than one value are candidates for color-coding. By default, the most covered candidate path is used. For instance, the default for collection C1 is \texttt{wdt:P31 instance of}, its summary is composed of two values: \texttt{wd:Q1004 Comics} and the aggregate \texttt{Other}. Entities are colored in blue for the former, in green for the latter, or with a gradient if they hold several values. Users can select another path to color the entities with the color button \wsv{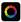} in the top bar. 
When a subset of entities is selected, they are colored in pink and others in black.

\subsection{Paths histograms}
Next to the map giving a visual overview of the entities, the histograms (\autoref{fig:teaser}, right) offer an overview of the aggregated completeness of each path, for the full set and the selected subset. Stacked charts embedded in the histograms give access to the distribution of each path. They are laid out in mirror to let users compare the profiles of the subset and the full set, in terms of completeness and distribution. They support the following tasks:
\begin{itemize}[nosep]
\item see and compare the homogeneity of the full set with the selected subset, regarding all properties
\item see and compare the completeness and distributions of the full set with the selected subset
\item select entities based on the presence or absence of a property
\item select entities based on summarised distributions of the values, languages, and datatypes at the end of the paths.
\end{itemize}

\subsubsection{Overview of the completeness}
The grey bars represent all paths describing the collection, ordered by completeness, to give another visual signature of the completeness, showing at first glimpse the number of paths fully complete. \autoref{fig:charts} shows paths summaries for the collections displayed in \autoref{fig:maps}. The map and the histogram are linked and coordinated.  

\begin{figure}
	\frame{\includegraphics[width=\columnwidth]{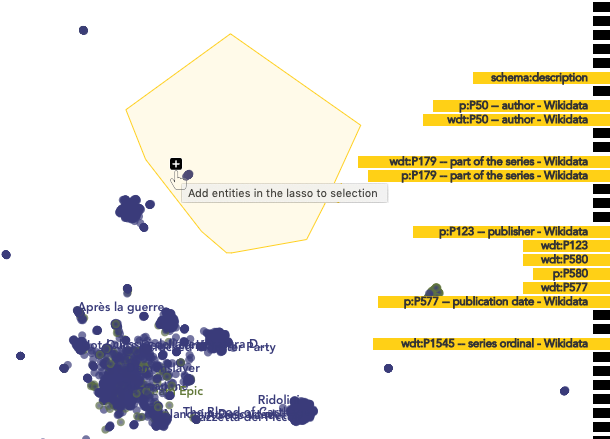}}
  	\caption{Hovering a predefined zone on the map highlights it in yellow, and gives access to the $+$ button, to use it as a condition for a selection. It also displays and highlights in yellow the names of the paths missing for the entities in this zone.}
	\label{fig:selectionzone}
\end{figure}

Each row represents a path; the length of the grey bar is mapped to its percentage of completeness. Clicking on a path opens it, showing a summary as detailed in the next paragraph.
Paths labels are displayed on the left of each row. By default, they appear when users hover a path, when they hover a predefined zone on the map, as in \autoref{fig:selectionzone}, or when a path is open. Users can toggle them on permanently as in \autoref{fig:teaser}with the labels button \wsv{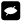}.

\subsubsection{Summarised distributions characterizing a path}\label{sec:summary:open}
When open, a path displays a summary of the distribution of the values at its end, as well as of their datatypes and languages. This work builds on the concept of \emph{semantic paths}. Originally, their description was limited to the counts of unique and total values at the end of the chain~\cite{destandauSPaths2020}. We extend them by adding a more detailed summary of their distribution. Our API takes as parameters \emph{the URI of a SPARQL endpoint}---a service accepting SPARQL query over an RDF dataset, \emph{a similarity criteria for the collection}, \emph{a maximum depth for the chains of properties} to analyse. We first retrieve all the path patterns---the combinations of chains of properties that will be analysed---up to the max depth, and their completeness rate. Then, to be able to compute summaries on any subset in a time that is acceptable for interaction, we retrieve the values at the end for all entities and store them in a matrix that can be processed rapidly with Python. We precompute a summary of the collection. The summaries of subsets will be computed on-demand. 


\begin{figure*}
  \centering
  \frame{\includegraphics[width=\linewidth]{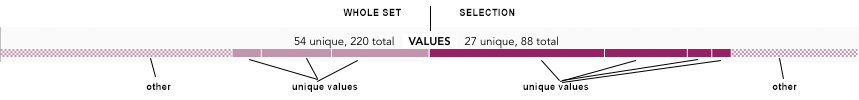}}
  \caption{Summary of values for a path: the whole set is presented on the left, in comparison to the selection on the right. The summary details values representing more than 5\% of the total, and aggregates others: for the whole set, only 3 of the 54 unique values are well represented enough to be detailed; the 51 that remain are merged in the `other' rectangle, represented with a dotted texture.  Hovering a rectangle displays the label and count of the value it represents. Each value, including the aggregate, can be clicked to be added as a condition for a selection.}
  \label{fig:comparesummary}
\end{figure*}

\begin{figure}
  	\includegraphics[width=\columnwidth]{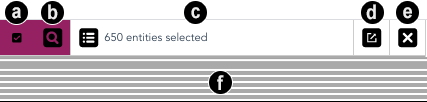}
  	\caption{The selection bar contains controls to inspect and refine the conditions for a selection and its result. The number of checkboxes in ( a) shows how many conditions are pending (here, there is one). Clicking on (a) displays the query in pseudo code (see \autoref{fig:constraints}). Clicking on (b) retrieves the list of entities matching the conditions and their summary. When a selection has been retrieved, (c) indicates the number of the list of entities in the selection, clicking on it displays the list in \autoref{fig:entities}. (d) enables to export the selection, and (e) to clear it.\bigskip\bigskip}
	\label{fig:selectiontools}
\end{figure}

\begin{figure}
  	\frame{\includegraphics[width=\columnwidth]{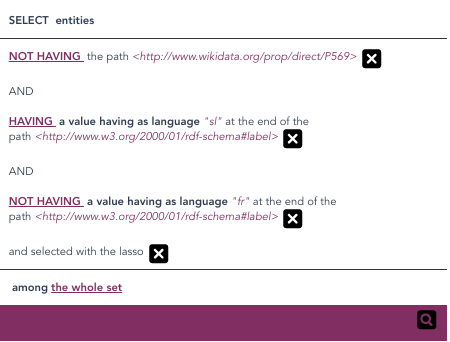}}
  	\caption{Conditions for a selection are expressed in pseudo code, to let users understand how the tool retrieves entities. They can refine them by toggling the elements that are underlined : `having' can be switched to `not having', resulting in the inverse condition, and `the whole set'' to `the current selection'.}
	\label{fig:constraints}
\end{figure}

\begin{figure}
  	\frame{\includegraphics[width=\columnwidth]{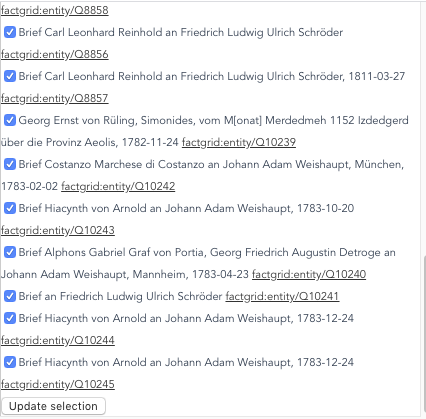}}
  	\caption{List of entities in the current selection. The label is in the preferred language when available. Clicking on the URI opens it in a new window.\bigskip\bigskip}
	\label{fig:entities}
\end{figure}

\subsubsection{Comparison of the full set with the selected subset}

To make sense of a \emph{subset} of entities, users need to identify its distinctive features, what defines it in comparison to the whole collection. The histogram is laid out as a mirror of the histogram for the full collection, to facilitate this comparison. 
For instance in \autoref{fig:teaser}, comparing the two histograms shows that the subset is very homogeneous; although it misses important information (no grey bar in the right column), the 16 paths that are described are complete (full grey bar in the right column), while only 8 of them are fully complete for the full set. Paths missing in the subset are highlighted in yellow, to help users focus on the problem they are trying to solve. 

To support users in the comparison task, the tool also draws their attention to which paths to inspect in order to understand the specificity of a selection (how it differs from the full set); it colors them in pink. 
\label{sec:distrib:interest}

The yellow color indicates paths that are missing in the subset. It stands out, more intense and luminous than the other colors in the interface, to draw the attention of users to what is not there, and help them make sense of the absence.

\subsection{Selection bar}\label{sec:constraints}
The selection bar supports users in inspecting and refining the conditions for a query. 
\emph{Conditions} are selection criteria in the database sense, combined by a conjunction (an ``and'' operator). Hovering over the map highlights predefined zones (\autoref{fig:selectionzone}). The $+$ button in the centre of the zone allows adding the zone as a condition. Clicking on the map switches from region to lasso mode, to let users select zones that are not predefined. Graphical elements in the histograms and the summaries can be added to and removed from the selection.
The selection control bar in \autoref{fig:selectiontools} supports users in understanding what happens when they add a condition, validating the selection, seeing the list of entities selected, and clearing the selection.

\begin{enumerate}[label=\alph*), nosep]
\item \emph{Toggle list of conditions}. Each condition is represented by a checked box.
When at least one condition has been added, (a) and (b) become pink, to indicate that the selection can be queried. Clicking (a) toggles the list of conditions, as shown in~\autoref{fig:constraints}. The query is written in pseudo-code; users can remove conditions from the list, toggle them to their inverse condition, or toggle the scope of the query from `whole collection' to `current subset'.
\item \emph{Inspect selection}. The combination of conditions defines the selection. When users clicked the inspect button, the query is sent to our Python API. The new list of entities in the selection is retrieved, and \autoref{fig:selectiontools}-c is updated first. Then the summary for the entities is computed and displayed under the selection control bar \autoref{fig:selectiontools}-f.
\item \emph{Toggle list of selected entities}. Clicking this button toggles the list in \autoref{fig:entities}. Users can remove entities from the list. Clicking the `Update selection' button at the bottom updates the paths summary for the selection.
\item \emph{Export selection}. This button triggers the download of 3 \texttt{csv} files that can be used to keep track of the query:
\texttt{condition.csv} contains the list of conditions used to get the selection,
\texttt{selection.csv} contains the list of entities in the selection (URI + label), and \texttt{summary.csv} contains the summaries for the subset and full set.
\item \emph{Clear selection}. Clears the current selection and its summary.
\end{enumerate}

\medskip


\section{Scenario of Use}

We designed our tool to help users see what is missing in their dataset and make sense of it. Let us describe the interface from the point of view of a contributor, Alice, who wants to curate Wikidata entities of class \texttt{Q1004 Comics}, describing comic books. She opens the tool, sees the map of entities in \autoref{fig:teaser}. As she moves the mouse, yellow zones delimiting clusters of entities appear, and paths that are missing for the zone are highlighted in yellow. Her attention gets caught by a small cluster, which misses many pieces of information that are important to describe comics, such as \texttt{P407 language of work or name}, \texttt{P495 country of origin}, \texttt{P123 publisher}, \texttt{P577 publication date} and  \texttt{P136 genre}. 
She decides to inspect this group in more detail: she adds this zone to the conditions for selection using the $+$ symbol and validates the selection with the magnifier button.
The selection bar announces a total of 20 entities, and the summary appears under it. 
Some of the paths are colored in pink, indicating that their summary for the selection might be significantly different from the full set.
Alice hovers the paths highlighted in pink to see their labels and starts by opening \texttt{rdfs:label}. 
She notices that there are 20 distinct labels, all of them in French. Then, she inspects \texttt{schema:description}. 
Its summary reveals that a single value is repeated 20 times: ``stripverhaal van Robbedoes en Kwabernoot'' (``comic strip Spirou \& Fantasio'' in Dutch, a popular comic strip originally written in French).
The 20 descriptions are in Dutch.
She inspects \texttt{schema:dateModified} and sees that 20 entities were last modified on the same day.
The \texttt{P179 part of the series} property indicates that 20 are part of the same series. 
Alice finds that those entities appear to have very similar needs. According to her quality standards, labels and descriptions should be available in similar languages (as opposed to labels being in French only and descriptions in Dutch only). From what she knows, Spirou and Fantasio comics are known enough that it should be easy to find the author, language, publisher, and publication date. 
The information can likely be found from the same sources for at least some of the albums. If Alice is lucky, one of the sources might even be the URI of the series that all entities belong to.
It looks like she will be able to save time by fixing those entities at once. Now that she has identified that this cluster needs a certain type of action, she would like to make sure that she will check all the entities belonging to the series, even if they miss slightly different information and are not in the initial cluster.
To do so, she clicks on the value shared by 20 entities to add it to conditions for selection.
She then opens the conditions and reads the query: ``SELECT  entities HAVING  the value \texttt{wd:Q1130014} at the end of the path \texttt{wdt:P179} among the current selection''. 
She toggles the scope definition from ``current selection'' to ``full set'' and validates the selection with the magnifier button. 
The selection bar now announces a total of 35 entities, all part of the ``Spirou and Fantasio'' series.  She clicks the export button and downloads the files describing this group for fixing it later.

She then hovers the next zone. The paths highlighted in yellow indicate that entities in this zone also miss similar important information, the main difference being that they have a \texttt{skos:altLabel}, but no attribute \texttt{wikibase:timeStamp}.
Note that even if the properties discriminating two neighbour zones do not appear to be meaningful properties, this structural approach helps detect coherent subsets.
In order to inspect the new cluster, she adds the zone to conditions for selection using the $+$ symbol and validates the selection with the magnifier button.
The new selection replaces the previous one. The selection bar announces 127 entities. 100\% of them have a \texttt{P179 part of the series}, so she opens the summary for this path that is now colored in pink, hoping that she can detect interesting groups.
The summary announces 25 unique values, and 3 values stand out because they are well represented. Those values are URIs, and she hovers them to dereference them in the URI bar above the map; she sees the corresponding labels: ``Sammy'' (25), ``Bobo'' (21), and ``Natacha'' (14). The rest of the values are merged in an `other' group (67). She clicks on the first value to add it to conditions for selection and validates the selection with the magnifier button. She exports this selection. She repeats the same actions with the two other subgroups. Now she can refer to the \texttt{csv} files she has exported to fix each of those 3 groups. 

This exploratory approach enables her to quickly detect small groups that are coherent and thus easy to fix. Let's now see how she can use the tool starting from the summary of paths. She clears the current selection and clicks on the eye pictogram to display all path labels. She figures out at first glance, from the length of the grey bars in the histogram, that less than half of the entities have an author. She decides to make this a priority to fix. She opens the author summary, which confirms a completeness rate of 42\%, and she clicks on the bar to add it to conditions for selection. She opens conditions to read the query: ``SELECT  entities HAVING  the path \texttt{wdt:P50} among the whole set''. She toggles the condition from `HAVING' to `NOT HAVING' and validates the magnifier button. The selection bar displays 1929 entities for the selection. The summaries for paths are mainly composed of `other' values. Wondering how to deal with this huge list, she considers refining the selection by combining conditions. She sees the property \texttt{P3589 Grand Comics Database Series ID} in the list. She decides to inspect entities having no author but such an identifier, which might mean that the information about the author will be accessible. The subset counts 49 entities, which is indeed more manageable. She exports the selection; the workflow should be easy since the source is the same for all entities; it might even be automatable. There are still 1880 entities without authors. She tries another strategy, looking for entities that have a publisher but no author. The result counts 129 entities. 

With The Missing Path, incompleteness can be explored starting from the map or from the summary and then switching between them to refine or expand the exploration.

\section{User Study: Iterative Design and Evaluation}

Using a methodology inspired by MILCS~\cite{Shneiderman2006} we worked with Wikidata contributors to validate our approach and iteratively improve the design of the tool. This methodology is optimised to evaluate creativity support tool, and analysing incompleteness is a task that demands creativity, with no established method or measure to assess its effectiveness.
It relies on an acute knowledge of the data and the workflow underlying their creation and edition.

\subsection{Participants}

We recruited 9 Wikidata contributors (2 female, 7 male) via calls on Wikidata mailing lists and Twitter. 3 were based in France, 1 in Sweden, 1 in Germany, 1 in the Netherlands, 1 in Australia, and 1 in the USA. 4 of them used Wikidata in the context of their work, and 5 as volunteers. They were 30 to 59 years old (avg: 39.89 yo, median: 34 yo). Their experience contributing ranged from 6 months to 7 years (avg: 3.46 years, median: 4 years). They spent between 1 and 165 hours a month contributing (avg: 52.89 hours, median: 24 hours). All participation was voluntary and without compensation.

\subsection{Set-up}
The interviews were lead online through a videoconferencing system. We used an online survey form to guide participants through the first interview and to collect demographic information. Our tool was run on a web server hosted by the laboratory and logs were filed in a database on our server.

\subsection{Procedure}
\subsubsection{First interview}
After going through the informed consent form and collecting demographic information, the interview was guided by the following question: 
\begin{em}
\begin{enumerate*}
\item Which Wikidata projects do you contribute to?
\item How do you decide which data you will update in priority? 
\item Did it ever happen that you wanted to contribute and didn't know where to start? 
\item Can you tell me about the last item you edited? 
\item Do you propose items for others to update? How do you select them?
\end{enumerate*}
\end{em}
Then we gave a quick overview of the tool and asked participants if they would be interested in visualising a collection with it. 

\subsubsection{Second interview}
We first shared our screen with participants to present the tool and its documentation. We demonstrated basic tasks on the Comics collection in a 5 minutes demo. 
Then participants took control, sharing their screen so that we were able to observe them. They registered their unique identifier in the tool for logs and performed the same tasks on their own collections.
We explained to them how to give feedback using Gitlab issues. These Issues can be of three types: feature, problem, and insight. 
We encouraged participants to use any other communication channel if they felt more comfortable with it, explaining that we would transform it into issues ourselves.
At the end of the interview, we created issues to file the reactions we have observed during the interview.

\subsubsection{Follow-up}
We communicated with participants by email (and a mix of Twitter direct messages and email for one of them). We conducted an additional video interview with four of them, during which we assisted them with the use of the tool when needed.

\medskip

We name our participants P1 to P9, according to their unique identifier.
We logged a total of 298 actions attributed to our participants, distributed as follows: add a condition (46), remove from condition (20), retrieve subset (74), compute projection (21),  clear selection (21), load collection (61), and selectColor (55). P1 had no logs at all --- his web browser privacy settings interfered with our log collection mechanism, although he reported using the tool.  
Over 4 months we conducted a total of 22 interviews, with an average of $2.44$ interviews per participant (median 3), and we received a total of 111 emails or Twitter direct messages, with an average of 12.33 messages per participant (median 11). We extracted a total of 78 issues. Only three were filed directly by a participant; we transcribed all others from the interviews (54) and emails (19). 
One participant dropped out after the first interview, and one after the second, without giving a reason.

We used a total of 12 collections during the study, as listed in \autoref{tab:collection}. 
Comics was our demo collection. Each participant had an initial collection, and three asked for the analysis of an additional collection during the process. The one who dropped out after the first interview had no collection.

\subsection{Data collection and analysis}
We recorded the first interview. For the second and third interviews, we relied on our notes to transcribe issues right after the interview. We also transcribed issues from emails and messages we received. 
At the end of the study, we exported the answers to the form and the issues into \texttt{csv} files, and we tagged the type (one of \textit{collection, feature, general comment, insight, problem}) and the status (one of \textit{solved, not relevant, future work}) of issues.

\subsection{Results}

We analyse the results with regards to \md{usability issues and validation of the approach.}

\begin{table}
	\centering
	\begin{tabular}{llcc}
		\textbf{ID} &\textbf{Description} & \makecell{\textbf{number of} \\\textbf{entities}}& \makecell{\textbf{number of} \\\textbf{paths}}\\
		\toprule
		C1 &Comics  & 4567 & 401 \\
		C2 &French deputies & 14513 & 1350 \\
		C3 &BFI movies & 6666 & 985 \\
		C4 &Ice Skating 1 & 2204 & 94 \\
		C5 &Ice Skating 2 & 1377 & 70 \\
		C6 &Illuminati* & 7938 & 183 \\
		C7 &Maps & 142 & 109  \\
		C8 &Monuments in France & 48845 & 775 \\
		C9 &Monuments in Brittany  & 4210 & 367 \\
		C10 & Research institutes & 235 &  353\\		
		C11 &Swedish female sculptors & 292 &  395\\
		C12 &Swedish photographers & 760  &  739\\
		\bottomrule
	\end{tabular}
	\caption{Data collections visualised in the tool for the evaluation, available in the demo instance. * The Illuminati collection comes from an instance of Wikibase, Factgrid}\label{tab:collection}
\end{table}

\subsubsection{\md{Usability issues}}

\md{The iterative design process helped us solve usability issues.}
The most critical issue was the understanding of the map. In an earlier version of the tool, the interface emphasised information missing in a subset after its summary was retrieved. P3 stated that he found it difficult to understand which paths were missing. We decided to \md{precompute} default zones on the map and to display missing path names on hover to make the interface self-explanatory. We also added the yellow color to highlight what was missing. After this, users reacted much more positively to the map: ``I understand now'' (P2), ``Now I understand it better'' (P3).

\md{A second issue was the difficulty to identify a distinctive feature in the selection.}
P7 suggested highlighting the paths for which the summary appears to be significantly different in the subset than in the full set. In an earlier version, inspecting a cluster to understand its specificity necessitated looking at each path one by one, which was long and uneasy. Participants did not know where to start, and it could happen that they repeatedly opened paths for which the summary consisted in `other' aggregates, which was not much help to identify the specificity of a group. We added the automatic detection of significant differences, as described in section \nameref{sec:distrib:interest}, and highlighted them in pink. 
This feature saves substantial time and provides guidance. 

\begin{figure}
  \frame{\includegraphics[width=\columnwidth]{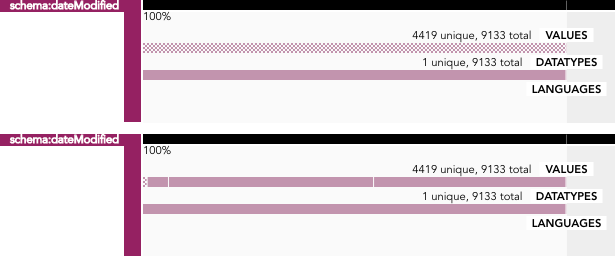}}
  	\caption{Evolution of the layout for dates summaries during the iterative process. This is the summary for the path \texttt{schema:dateModified} on the collection \texttt{C1 Comics}. In the first version (top) the dates were grouped by unique values, which very often resulted in an `other' aggregate, laid out with a dotted texture. After participants' feedback we implemented binning for dates  (bottom), which results in 4 groups, from right to left: ``2018'' (4150), ``2019'' (4423), ``2020'' (460), and `other' (100) --- hovering the rectangles reveal the value and counts. Each value can be used as a condition for selection.}
	\label{fig:dates}
\end{figure}

The way we presented summaries also evolved during the process. We had first designed summaries for integers as boxplots, thinking it could be interesting for users to select only outliers or median. We realised that our users could not read boxplots and ignored those summaries, so we switched to a stacked chart of unique values, similar to the one used for text values. On the other hand, dates and times were initially designed as a stacked chart, and most of the time resulted in a single `Other' aggregate. P1 asked if we could group dates, so we implemented binning into hours, days, months, or years. This feature improved the usability of some path summaries like, for instance, for instance, the modification date, as we can see in \autoref{fig:dates}. 
When he first used the tool, P1 also tried to select the `other' aggregate as a condition for selection, which was at the time not possible. We also added this feature.

\md{All in all, participants suggested 32 new features and reported 15 problems. We implemented 20 of the new features, marked 3 as irrelevant in the context of our work, and kept 9 for future work. We solved 13 problems, marked one as an exception, and one for future work.}

\subsubsection{Validation of the approach}

We were particularly interested in knowing if users would rely on it to start the exploration of subsets. P1, P9, and P2 did. 
P1 explained: ``I see it as a way to start the exploration, see the outlines''. He had already spent a lot of time curating this set of data and knew them well. However, there are more than 14,000 entities in the set, and he worked more specifically on those related to the French Fifth Republic, so the map was useful to spot problems he was not aware of. For instance, the first cluster he inspected during the second interview was a set of 47 deputies having no place of birth. He commented: ``There should not be entities with no place of birth. This group can easily be fixed, the information is available through the Sycomore French deputies database, and they all have a Sycomore ID'' (\texttt{wdt:P Sycomore ID}). 
During the third interview, another cluster showed entities (deputies) with no given name. He explained: ``All deputies should have a given name. This can be fixed easily from the labels.'' He thought that even if the focus might switch from the map to the histogram as you get to know your data better and they become more homogeneous, there can always be new stages when you incorporate new sets of entities and want to bring them to the same level of quality as the rest of the data when the map could prove to be useful again.

P9 did also start from the map. He was planning to import and manage his own catalogue of movies in Wikidata. Since he was still at a  planning step, we had selected the BFI movie database, which was about similar in size and type of information to what his own data would later be. He figured out there was a cluster of 16 entities without titles. He inspected the summary and found out those entities all had a label, which meant the titles would be very easy to fix. A double-check through the histogram showed that there were 125 entities with no title but a label. Another cluster had no directors. This leads him to use the histogram to look for all entities having no directors, which amounted to 1380 entities. Looking at the map, he could see they were spread into about 20 different clusters, depending on what else was missing. Hovering the clusters then gave him an overview of the possible combination of missing attributes. He inspected two of them in more detail. Trying to imagine how he could use the tool later with his own data, he said he would probably want to configure the projection with paths he wished to achieve full completeness for, and then work on the data until there's only one big cluster.

P2 needed to customize the map, using only the paths that were of prior importance for him to compute the projection. This reduced the map to a few clusters that he found meaningful. ``Now I am satisfied. This is the image I wanted when all the irrelevant criteria that complexified the map have been removed.'' Then he started his exploration from the histogram.  He used the combination of conditions to find the list of all monuments qualified as churches --- having \texttt{wd:Q16970 church building} as a value for \texttt{wdt:P31 instance of} --- but with no identifier \texttt{wdt:P3963 Clochers de France ID}, specific to churches. He expressed the wish to see the entities highlighted on the map, a feature described in section \nameref{sec:map:color}, that we added following his demand. While explaining that \texttt{wdt:P18 images} was not a relevant path for the projection in his opinion, because it was normal that some entities had no images, he exclaimed ``I know what I am going to do this afternoon!'' He had figured out he could select all the entities having no \texttt{wdt:P18 image} but a \texttt{wdt:P373 Commons category}, because if they had a Commons identifier, then he knew he could find an image. He added ``I could have done the same with SPARQL, but I would never have had the idea. The tool gave me the idea.'' 

P5 preferred to start from the summaries and ignored the map. 
\md{She suggested a feature to support better exploration from the summaries: the possibility to combine conditions to refine the selection. Interestingly, this made our tool much more flexible, able to support more diverse tasks.}


\begin{figure}
  \frame{\includegraphics[width=\columnwidth]{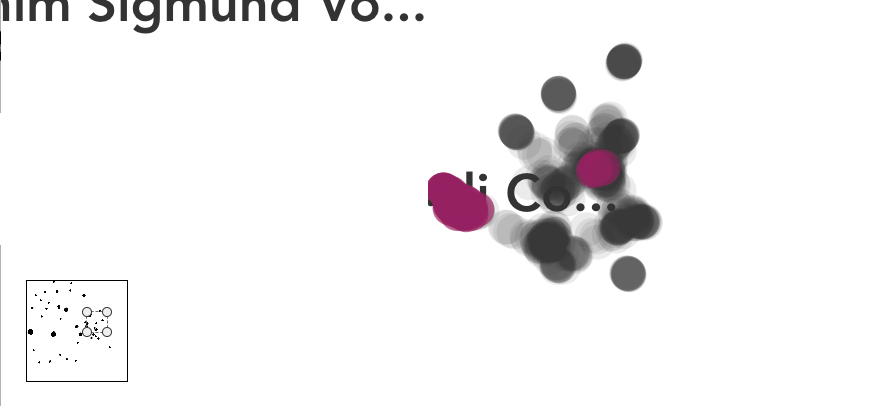}}
  	\caption{Entities highlighted on the map of the collection C6 when all entities having a \texttt{factgrid:prop/P17 Dataset complaint} are selected. The contributor who made those statements explained he worked on small groups of consistent entities, and we can see they appear as such on our map, although P17 is not used to compute the map. This shows that those consistent groups miss the same well represented attributes.}
	\label{fig:complaint}
\end{figure}

\md{In total, participants made 16 general comments on the approach and reported 12 insights on their data.}

\medskip

In summary, our study helped us make the tool more flexible and adapt it to different workflows.  
We had first thought the map would be the main entry point, and statistical summaries would help refine and analyse the clusters. We realised that looking at the histogram overview did also trigger ideas of specific completeness profiles (e.g.\ entities missing a specific path but not missing another one, or entities with a specific feature and missing a path), which is another way to detect coherent clusters. The full list laid flat triggered associations that could be quickly verified. 

\section{Conclusion and Future Work}
We have presented \TMP, a visualisation tool to support data producers in analysing incompleteness in their data to identify subsets of items that can be fixed. It is based on two novel representations of RDF data; the map provides a structural snapshot of a collection, reflecting its history and allowing users to untangle its various strata; the histograms and stacked charts laid out in mirror allow comparing a subset with the full collection, revealing its distinctive features. The coordination of those new visualisations supports users in the interactive exploration and analysis of incomplete subsets.
Our user study confirmed that Wikidata contributors could gain new insights and identify groups of entities that can be fixed. Participants guided us to make the tool more understandable and usable. Doing so, they also lead us to make it more flexible, supporting various workflows, and this pushed our tool in the direction of an exploratory analysis tool. 

To our knowledge, there is no such tool for RDF data. In the future, we would like to investigate other analysis scenarios, besides incompleteness. We will also address the need to keep track of the exploration in the tool, not only by exporting the data so that users can monitor the evolution of their collection.

Having heard of our tool, Wikidata product managers became intrigued, interested, and asked for a demonstration. As one of them told us when we demonstrated the tool, ``One of the big problems our contributors face in keeping the data quality and completeness high is the fact that it is very hard to see the big picture due to Wikidata's modelling being centred around individual entities. Your tool is addressing this issue''. We will continue to interact with the Wikidata community and other RDF data producers to improve our tool and support better quality Knowledge Graphs.

\begin{acks}
The authors wish to thank all the participants in the experiment.
\end{acks}

\bibliographystyle{SageV}

\bibliography{main}

\end{document}